\documentclass[unsortedaddress,nopreprint,twocolumn,numbers]{jasatex}

\usepackage{amssymb, latexsym}
\usepackage{amsmath} 
\usepackage{amsthm}
\usepackage{bm}
\usepackage{graphicx}
\usepackage{epstopdf}  
\usepackage[mathscr]{eucal}
\usepackage{enumerate}

\graphicspath{{./Figures/}{./}}

\newcommand{\be}{\begin{equation}}
\newcommand{\ee}{\end{equation}}

\begin{document}

\title[Dictionary learning of sound speed profiles]{Dictionary learning of sound speed profiles}

\author{Michael Bianco}
\email[Corresponding author. Electronic mail: ]{mbianco@ucsd.edu}
\author{Peter Gerstoft}
\affiliation{Scripps Institution of Oceanography, University of California San Diego, La Jolla, California 92093--0238}

\date{\today}

\begin{abstract}
To provide constraints on their inversion, ocean sound speed profiles (SSPs) often are modeled using empirical orthogonal functions (EOFs). However, this regularization, which uses the leading order EOFs with a minimum-energy constraint on their coefficients, often yields low resolution SSP estimates. In this paper, it is shown that dictionary learning, a form of unsupervised machine learning, can improve SSP resolution by generating a dictionary of shape functions for sparse processing (e.g. compressive sensing) that optimally compress SSPs; both minimizing the reconstruction error and the number of coefficients. These learned dictionaries (LDs) are not constrained to be orthogonal and thus, fit the given signals such that each signal example is approximated using few LD entries. Here, LDs describing SSP observations from the HF-97 experiment and the South China Sea are generated using the K-SVD algorithm. These LDs better explain SSP variability and require fewer coefficients than EOFs, describing much of the variability with one coefficient. Thus, LDs improve the resolution of SSP estimates with negligible computational burden. \\
\copyright~2016 Acoustical Society of America 
\end{abstract}

\maketitle
\section{\label{sec:intro}Introduction}
Inversion for ocean sound speed profiles (SSPs) using acoustic data is a non-linear and highly underdetermined problem.\cite{gerstoft94} To ensure physically realistic solutions while moderating the size of the parameter search, SSP inversion has often been regularized by modeling SSP as the sum of leading order empirical orthogonal functions (EOFs).\cite{leblanc80}\textsuperscript{--}\cite{huang08} However, regularization using EOFs often yields low resolution estimates of ocean SSPs, which can be highly variable with fine scale fluctuations. In this paper, it is shown that the resolution of SSP estimates are improved using dictionary learning,\cite{rubenstein2010}\textsuperscript{--}\cite{engan2000} a form of unsupervised machine learning, to generate a dictionary of regularizing shape functions from SSP data for parsimonious representation of SSPs. 

Many signals, including natural images\cite{hyvarinen2009}\textsuperscript{,}\cite{jpeg2000}, audio\cite{gersho1991}, and seismic profiles\cite{taylor79} are well approximated using sparse (few) coefficients, provided a dictionary of shape functions exist under which their representation is sparse. Given a $K$-dimensional signal, a dictionary is defined as a set of $N$, $\ell_2$-normalized  vectors which describe the signal using few coefficients. The sparse processor is then an $\ell_2$-norm cost function with an $\ell_0$-norm penalty on the number of non-zero coefficients. Signal sparsity is exploited for a number of purposes including signal compression and denoising.\cite{elad2010} Applications of compressive sensing,\cite{candes06} one approximation to the $\ell_0$-norm sparse processor, have in ocean acoustics shown improvements in beamforming,\cite{edel11}\textsuperscript{--}\cite{choo16} geoacoustic inversion,\cite{yardim14} and estimation of ocean SSPs.\cite{bianco16}

Dictionaries that approximate a given class of signals using few coefficients can be designed using dictionary learning.\cite{elad2010} Dictionaries can be generated ad-hoc from common shape functions such as wavelets or curvelets, however extensive analysis is required to find an optimal set of prescribed shape functions. Dictionary learning proposes a more direct approach: given enough signal examples for a given signal class, learn a dictionary of shape functions that approximate signals within the class using few coefficients. These learned dictionaries (LDs) have improved compression and denoising results for image and video data over ad-hoc dictionaries.\cite{elad2010,schnass14} Dictionary learning has been applied to denoising problems in seismics \cite{beckouche14} and ocean acoustics \cite{taroudakis15,wang16}, as well as to structural acoustic health monitoring.\cite{alguri16}

The K-SVD algorithm,\cite{aharon06} a popular dictionary learning method, finds a dictionary of vectors that optimally partition the data from the training set such that the few dictionary vectors describe each data example.  Relative to EOFs which are derived using principal component analysis (PCA),\cite{hannachi2007,monahan2009} these LDs are not constrained to be orthogonal. Thus LDs provide potentially better signal compression because the vectors are on average, nearer to the signal examples (see Fig.\ \ref{fig:featureSpace}).\cite{engan2000} 

In this paper, LDs describing 1D ocean SSP data from the HF-97 experiment\cite{carbone2000}\textsuperscript{,}\cite{hodgkiss2002} and from the South China Sea (SCS)\cite{pinkel} are generated using the K-SVD algorithm and the reconstruction performance is evaluated against EOF methods. In Section II, EOFs, sparse reconstruction methods, and compression are introduced. In Section III, the K-SVD dictionary learning algorithm is explained. In Section IV, SSP reconstruction results are given for LDs and EOFs. It is shown that each shape function within the resulting LDs explain more SSP variability than the leading order EOFs trained on the same data. Further, it is demonstrated that SSPs can be reconstructed up to acceptable error using as few as one non-zero coefficient. This compression can improve the resolution of ocean SSP estimates with negligible computational burden.

\textit{Notation}: In the following, vectors are represented by bold lower-case letters and matrices by bold uppercase letters. The $\ell_p$-norm of the vector $\mathbf{x}\in\mathbb{R}^{N}$ is defined as $\|\mathbf{x}\|_p=\big(\sum^N_{n=1}\big|x_n\big|^p\big)^{1/p}$. Using similar notation, the $\ell_0$-norm is defined as $\|\mathbf{x}\|_0=\sum^N_{n=1}\big|x_n\big|^0=\sum^N_{n=1}1_{|x_n|>0}$. The $\ell_p$-norm of the matrix $\mathbf{A}\in\mathbb{R}^{K\times M}$ is defined as $\|\mathbf{A}\|_p=\big(\sum^M_{m=1}\sum^K_{k=1}\big|a_k^m\big|^p\big)^{1/p}$. The Frobenius norm ($\ell_2$-norm) of the matrix $\mathbf{A}$ is written as $\|\mathbf{A}\|_\mathcal{F}$. The hat symbol \ $\widehat{}$ \ appearing above vectors and matrices indicates approximations to the true signals or coefficients.

\section{EOFs and compression}
\subsection{EOFs and PCA}
Empirical orthogonal function (EOF) analysis seeks to reduce the dimension of continuously sampled space-time fields by finding spatial patterns which explain much of the variance of the process. These spatial patterns or EOFs correspond to the principal components, from principal component analysis (PCA), of the temporally varying field. \cite{hannachi2007} Here, the field is a collection of zero-mean ocean SSP anomaly vectors $\mathbf{Y}=[\mathbf{y}_1,...,\mathbf{y}_M]\in\mathbb{R}^{K\times M}$, which are sampled over $K$ discrete points in depth and $M$ instants in time. The mean value of the $M$ original observations are subtracted to obtain $\mathbf{Y}$. The variance of the SSP anomaly at each depth sample $k$, $\sigma_k^2$, is defined as
%
\begin{equation}
\sigma_k^2=\frac{1}{M}\sum_{m=1}^M \big(y_m^k\big)^2
\label{eq:sspMeanSub2}
\end{equation}
%
where $[y_1^k,...,y_M^k]$ are the SSP anomaly values at depth sample $k$ for $M$ time samples.

The singular value decomposition (SVD)\cite{hastie2009} finds the EOFs as the eigenvectors of $\mathbf{Y}\mathbf{Y}^{\rm{T}}$ by 
%
\begin{equation}
\mathbf{Y}\mathbf{Y}^{\rm{T}}=\mathbf{P}\mathbf{\Lambda}^2\mathbf{P}^{\rm{T}},
\label{eq:pcaAnal}
\end{equation}
%
where $\mathbf{P}=[\mathbf{p}_1,...,\mathbf{p}_L]\in\mathbb{R}^{K\times L}$ are EOFs (eigenvectors) and $\mathbf{\Lambda}^2=\rm{diag}([\lambda_1^2,...,\lambda_L^2])\in\mathbb{R}^{L\times L}$ are the total variances of the data along the principal directions defined by the EOFs $\mathbf{p}_l$ with
%
\begin{equation}
\sum_{k=1}^K\sigma_k^2=\frac{1}{M}\rm{tr}\big{(}\mathbf{\Lambda}^2\big{)}.
\label{eq:blah}
\end{equation}
%
The EOFs $\mathbf{p}_l$ with $\lambda_1^2\ge ... \ge \lambda_L^2$ are spatial features of the SSPs which explain the greatest variance of $\mathbf{Y}$. If the number of training vectors $M\ge K$, $L=K$ and $[\mathbf{p}_1,...,\mathbf{p}_L]$ form a basis in $\mathbb{R}^{K}$.

\subsection{SSP reconstruction using EOFs}
Since the leading-order EOFs often explain much of the variance in $\mathbf{Y}$, the representation of anomalies $\mathbf{y}_m$ can be compressed by retaining only the leading order EOFs $P<L$
%
\begin{equation}
\widehat{\mathbf{y}}_m=\mathbf{Q}_P\widehat{\mathbf{x}}_{P, m}
\label{eq:pcaAnal222}
\end{equation}
%
where $\mathbf{Q}_P\in\mathbb{R}^{K\times P}$ is here the dictionary containing the $P$ leading-order EOFs and $\widehat{\mathbf{x}}_{P, m}\in\mathbb{R}^{P}$ is the coefficient vector. Since the entries in $\mathbf{Q}_P$ are orthonormal, the coefficients are solved by
%
\begin{equation}
\widehat{\mathbf{x}}_{P,m}=\mathbf{Q}_P^{\rm{T}}\mathbf{y}_m.
\label{eq:eofPseudo}
\end{equation}
%
For ocean SSPs, usually no more than $P= 5$ EOF coefficients have been used to reconstruct ocean SSPs.\cite{huang08}\textsuperscript{,}\cite{gerstoft96} 
 
\subsection{Sparse reconstruction}
A signal $\mathbf{y}_m$, whose model is sparse in the dictionary $\mathbf{Q}_N=[\mathbf{q}_1 ,...,\mathbf{q}_N]\in\mathbb{R}^{K\times N}$ ($N$-entry sparsifying dictionary for $\mathbf{Y}$), is reconstructed to acceptable error using $T\ll K$ vectors $\mathbf{q}_n$.\cite{elad2010} The problem of estimating few coefficients in $\mathbf{x}_m$ for reconstruction of $\mathbf{y}_m$ can be phrased using the canonical sparse processor
%
\begin{equation}
\widehat{\mathbf{x}}_{m}=\underset{\mathbf{x}_m\in\mathbb{R}^N}{\arg\min} \|\mathbf{y}_m-\mathbf{Q}\mathbf{x}_m\|_2 \ \ \text{subject to} \ \ \|\mathbf{x}_m\|_0\le T.
\label{eq:sparseObject}
\end{equation}
%
The $\ell_0$-norm penalizes the number of non-zero coefficients in the solution to a typical $\ell_2$-norm cost function. The $\ell_0$-norm constraint is non-convex and imposes combinatorial search for the exact solution to Eq.\ (\ref{eq:sparseObject}). Since exhaustive search generally requires a prohibitive number of computations, approximate solution methods such as matching pursuit (MP) and basis pursuit (BP) are preferred.\cite{elad2010} In this paper, orthogonal matching pursuit (OMP)\cite{pati93} is used as the sparse solver. For small $T$, OMP achieves similar reconstruction accuracy relative to BP methods, but with much greater speed.\cite{elad2010}

It has been shown that non-orthogonal, overcomplete dictionaries $\mathbf{Q}_N$ with $N>K$ (complete, $N=K$) can be designed to minimize both error and number of non-zero coefficients $T$, and thus provide greater compression over orthogonal dictionaries.\cite{gersho1991}\textsuperscript{,}\cite{engan2000}\textsuperscript{,}\cite{elad2010} While overcomplete dictionaries can be designed by concatenating ortho-bases of wavelets or Fourier shape functions, better compression is often achieved by adapting the dictionary to the data under analysis using dictionary learning techniques.\cite{aharon06,engan2000} Since Eq.\ (\ref{eq:sparseObject}) promotes sparse solutions, it provides criteria for the design of dictionary $\mathbf{Q}$ for adequate reconstruction of $\mathbf{y}_m$ with a minimum number of non-zero coefficients. Rewriting Eq.(7) with
%
\begin{equation}
\underset{\mathbf{Q}}{\min}\big\{\underset{\mathbf{X}}{\min} \|\mathbf{Y}-\mathbf{Q}\mathbf{X}\|^2_\mathcal{F} \ \ \text{subject to} \ \ \forall_m,\|\mathbf{x}_m\|_0\leq T\big\},
\label{eq:dLearnObjective}
\end{equation}
%
where $\mathbf{X=[x}_1 ,...,\mathbf{x}_M]$ is the matrix of coefficient vectors corresponding to examples $\mathbf{Y}=[\mathbf{y}_1 ,...,\mathbf{y}_M]$, reconstruction error is minimized relative to the dictionary $\mathbf{Q}$ as well as relative to the sparse coefficients.

\begin{figure}
  \includegraphics[width = 2.7in]{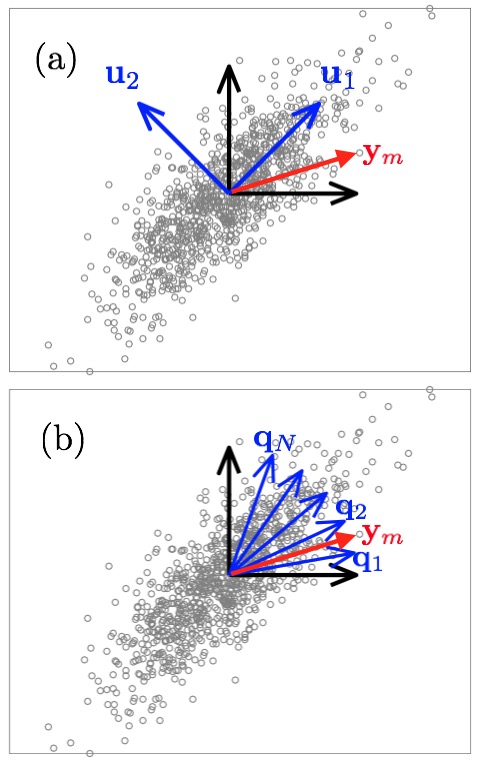}
\caption{(Color online) (a) EOF vectors $[\mathbf{u}_1,\mathbf{u}_2]$ and (b) overcomplete LD vectors $[\mathbf{q}_1 ,...,\mathbf{q}_N]$ for arbitrary 2D gaussian distribution relative to arbitrary 2D data observation $\mathbf{y}_m$.}
\label{fig:featureSpace}
\end{figure}

In this paper, the K-SVD algorithm, a clustering based dictionary learning method, is used to solve Eq.(\ref{eq:dLearnObjective}). The K-SVD is an adaptation of the K-means algorithm for vector quantization (VQ) codebook design (a.k.a. the generalized Lloyd algorithm).\cite{gersho1991} The learned dictionary (LD) vectors $\mathbf{q}_n$ from this technique partition the feature space of the data rather than $\mathbb{R}^{K}$, increasing the likelihood that $\mathbf{y}_m$ is as a linear combination of few vectors $\mathbf{q}_n$ in the solution to Eq.\ (\ref{eq:sparseObject}) (see Fig.\ \ref{fig:featureSpace}). By increasing the number of vectors $N\ge K$ for overcomplete dictionaries, and thus the number of partitions in feature space, the sparsity of the solutions can be increased further.\cite{engan2000}

\begin{figure}
  \includegraphics[width = 2.9in]{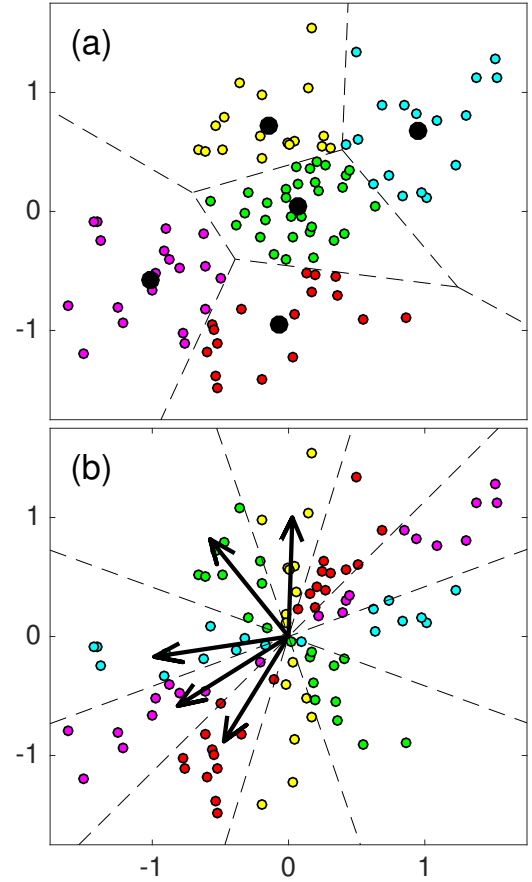}
\caption{(Color online) Partitioning of gaussian random distribution ($\sigma_1=0.75,\sigma_2=0.5$) using (a) 5 codebook vectors (K-means, VQ) and with (b) 5 dictionary vectors from dictionary learning (K-SVD, $T=1$).}
\label{fig:kmeans_vs_dlearn}
\end{figure}

\subsection{Vector quantization}
\noindent
Vector quantization (VQ)\cite{gersho1991} compresses a class of $K$--dimensional signals $\mathbf{Y}=[\mathbf{y}_1,..., \mathbf{y}_M]\in\mathbb{R}^{K\times M}$ by optimally mapping $\mathbf{y}_m$ to a set of code vectors $\mathbf{C}=[\mathbf{c}_1,..., \mathbf{c}_N]\in\mathbb{R}^{K\times N}$ for $N<M$, called a codebook. The signals $\mathbf{y}_m$ are then quantized or replaced by the best code vector choice from $\mathbf{C}$.\cite{gersho1991}  The mapping that minimizes mean squared error (MSE) in reconstruction
%
\begin{equation}
\rm{MSE}(\mathbf{Y},\widehat{\mathbf{Y}})=\frac{1}{N}\|\mathbf{Y}-\widehat{\mathbf{Y}}\|_\mathcal{F}^2,
\label{eq:distortion}
\end{equation}
%
where $\widehat{\mathbf{Y}}=[\widehat{\mathbf{y}}_1,..., \widehat{\mathbf{y}}_M]$ is the vector quantized $\mathbf{Y}$, is the assignment of each vector $\mathbf{y}_m$ to the code vectors $\mathbf{c}_n$ based on minimum $\ell_2$--distance (nearest neighbor metric). Thus the $\ell_2$--distances from the code vectors $\mathbf{c}_n$ define a set of partitions $(R_1,..., R_N)\in\mathbb{R}^K$ (called Voronoi cells) 
%
\begin{equation}
R_n=\left\{i\mid\forall_{l\neq n},\|\mathbf{y}_i-\mathbf{c}_n\|_2<\|\mathbf{y}_i-\mathbf{c}_l\|_2\}\right .,
\label{eq:clustering}
\end{equation}
%
where if $\mathbf{y}_i$ falls within the cell $R_n$, $\widehat{\mathbf{y}}_i$ is $\mathbf{c}_n$. These cells are shown in Fig.\ \ref{fig:kmeans_vs_dlearn}(a). This is stated formally by defining a selector function $S_n$ as
%
\begin{equation}
S_n(\mathbf{y}_m)=\bigg\{
\begin{matrix}
\ \ 1\ \ \text{if}\ \mathbf{y}_m\in\mathit{R_n} \\
\ 0\ \ \text{otherwise}.
\end{matrix}
\label{eq:selector}
\end{equation}
%
The vector quantization step is then 
%
\begin{equation}
\widehat{\mathbf{y}}_m=\sum_{n=1}^N S_n(\mathbf{y}_m)\mathbf{c}_n.
\label{eq:quantize}
\end{equation}
%

The operations in Eq.\ (\ref{eq:clustering}--\ref{eq:selector}) are analogous to solving the sparse minimization problem
%
\begin{equation}
\widehat{\mathbf{x}}_{m}=\underset{\mathbf{x}_m\in\mathbb{R}^N}{\arg\min} \|\mathbf{y}_m-\mathbf{C}\mathbf{x}_m\|_2 \ \ \text{subject to} \ \ \|\mathbf{x}_m\|_0= 1,
\label{eq:altCluster}
\end{equation}
%
where the non-zero coefficients $x_m^n=1$. In this problem, selection of the coefficient in $\mathbf{x}_m$ corresponds to mapping the observation vector $\mathbf{y}_m$ to $\mathbf{c}_n$, similar to the selector function $S_n$. The vector quantized $\mathbf{y}_m$ is thus written, alternately from Eq.\ (\ref{eq:quantize}), as
%
\begin{equation}
\widehat{\mathbf{y}}_m=\mathbf{C}\widehat{\mathbf{x}}_m.
\label{eq:altQuantize}
\end{equation}
\subsection{K-means}
Given the MSE metric (Eq.\ (\ref{eq:distortion})), VQ codebook vectors $[\mathbf{c}_1,..., \mathbf{c}_N]$ which correspond to the centroids of the data $\mathbf{Y}$ within $(R_1,..., R_N)$ minimize the reconstruction error. The assignment of $\mathbf{c}_n$ as the centroid of $\mathbf{y}_j\in R_n$ is
%
\begin{equation}
\mathbf{c}_n=\frac{1}{|R_n|}\sum_{j\in R_n}\mathbf{y}_j,
\label{eq:centroid}
\end{equation}
%
where $|R_n|$ is the number of vectors $\mathbf{y}_j\in R_n$. 

The K-means algorithm shown in Table \ref{algo:kmeans}, iteratively updates $\mathbf{C}$ using the centroid condition Eq.\ (\ref{eq:centroid}) and the $\ell_2$ nearest--neighbor criteria Eq.\ (\ref{eq:clustering}) to optimize the code vectors for VQ. The algorithm requires an initial codebook $\mathbf{C}^0$. For example, $\mathbf{C}^0$ can be $N$ random vectors in $\mathbb{R}^K$ or selected observations from the training set $\mathbf{Y}$. The K-means algorithm is guaranteed to improve or leave unchanged the $\rm{MSE}$ distortion after each iteration and converges to a local minimum.\cite{gersho1991}\textsuperscript{,}\cite{aharon06}

\begin{table}[h!]
\begin{center}
\caption{The K-means algorithm (Ref. \ \onlinecite{gersho1991}.)}
\label{algo:kmeans}
\begin{tabular}{ll}
\hline\hline
    & Given: Training vectors $\mathbf{Y}=[\mathbf{y}_1,...,\mathbf{y}_M]\in\mathbb{R}^{K\times M}$ \\ \hline
    & Initialize: index $i = 0$, codebook $\mathbf{C}^{0}=[\mathbf{c}_1^0,...,\mathbf{c}_N^0]\in\mathbb{R}^{K\times N}$, \\
    & $\rm{MSE}^0$ solving Eq.\ (\ref{eq:distortion})--(\ref{eq:quantize})\\
I: & Update codebook \\
       & \hspace{2ex} 1. Partition $\mathbf{Y}$ into $N$ regions $(R_1,..., R_N)$ by \\
        & \hspace{4ex} $R_n=\left\{i\mid\forall_{l\neq n},\|\mathbf{y}_i-\mathbf{c}_n^i\|_2<\|\mathbf{y}_i-\mathbf{c}_l^i\|_2\}\right.$ (Eq.\ (\ref{eq:clustering}))\\
    & \hspace{2ex} 2. Make code vectors centroids of $\mathbf{y}_j$ in partitions $R_n$ \\
       & \hspace{4ex} $\mathbf{c}_n^{i+1}=\frac{1}{|R_n^i|}\sum_{j\in R_n^i}\mathbf{y}_j$ \\
 II. & Check error   \\
 & \hspace{2ex} 1. Calculate  $\rm{MSE}^{i+1}$ from updated codebook $\mathbf{C}^{i+1}$\\
 & \hspace{2ex} 2. If $|\rm{MSE}^{i+1}-\rm{MSE}^{i}|<\eta$ \\
 & \hspace{6ex} $i=i+1$, return to I \\
 & \hspace{5ex} else \\
       & end \\
\hline\hline
\end{tabular}
\end{center}
\end{table}
\section{Dictionary learning}
Two popular algorithms for dictionary learning, the method of optimal directions (MOD)\cite{engan2000} and the K-SVD,\cite{aharon06} are inspired by the iterative K-means codebook updates for VQ (Table \ref{algo:kmeans}). The $N$ columns of the dictionary $\mathbf{Q}$, like the entries in codebook $\mathbf{C}$, correspond to partitions in $\mathbb{R}^K$. However, they are constrained to have unit $\ell_2$-norm and thus separate the magnitude (coefficients $\mathbf{x}_n$) from the shapes (dictionary entries $\mathbf{q}_n$) for the sparse processing objective Eq.(\ref{eq:sparseObject}).
When $T=1$, the $\ell_2$-norm in Eq.\ (\ref{eq:sparseObject}) is minimized by the dictionary entry $\mathbf{q}_n$ that has the greatest inner product with example $\mathbf{y}_m$.\cite{elad2010}  Thus for $T=1$, $[\mathbf{q}_1,..., \mathbf{q}_N]$ define radial partitions of $\mathbb{R}^K$. These partitions are shown in Fig.\ \ref{fig:kmeans_vs_dlearn}(b) for a hypothetical 2D ($K=2$) random data set. This corresponds to a special case of VQ, called gain-shape VQ.\cite{gersho1991} However, for sparse processing, only the shapes of the signals are quantized. The gains, which are the coefficients $\mathbf{x}_m$, are solved. For $T>1$, the sparse solution is analogous to VQ, assigning examples $\mathbf{y}_m$ to dictionary entries in $\mathbf{Q}$ for up to $T$ non-zero coefficients in $\mathbf{x}_m$. 

Given these relationships between sparse processing with dictionaries and VQ, the MOD\cite{engan2000} and K-SVD\cite{aharon06} algorithms attempt to generalize the K-means algorithm to optimization of dictionaries for sparse processing for $T\ge1$. They are two-step algorithms which reflect the two update steps in the K-means codebook optimization: (1) partition data $\mathbf{Y}$ into regions $(R_1,..., R_N)$ corresponding to $\mathbf{c}_n$ and (2) update $\mathbf{c}_n$ to centroid of examples $\mathbf{y}_m\in R_N$. The K-means algorithm is generalized to the dictionary learning problem Eq.(\ref{eq:dLearnObjective}) as two steps:
\begin{enumerate}
  \item  Sparse coding: Given dictionary $\mathbf{Q}$, solve for up to $T$ non-zero coefficients in $\mathbf{x}_m$ corresponding to examples $\mathbf{y}_m$ for $m=[1,...,M]$
  \item Dictionary update: Given coefficients $\mathbf{X}$, solve for $\mathbf{Q}$ which minimizes reconstruction error for $\mathbf{Y}$.
  \end{enumerate}

The sparse coding step (1), which is the same for both MOD and K-SVD, is accomplished using any sparse solution method, including matching pursuit and basis pursuit. The algorithms differ in the dictionary update step.

\subsection{The K-SVD Algorithm} 
The K-SVD algorithm is here chosen for its computational efficiency, speed, and convergence to local minima (at least for $T=1$). The K-SVD algorithm sequentially optimizes the dictionary entries $\mathbf{q}_n$ and coefficients $\mathbf{x}_m$ for each update step using the SVD, and thus also avoids the matrix inverse. For $T=1$, the sequential updates of the K-SVD provide optimal dictionary updates for gain-shape VQ.\cite{aharon06}\textsuperscript{,}\cite{gersho1991} Optimal updates to the gain-shape dictionary will, like K-means updates, either improve or leave unchanged the MSE and convergence to a local minimum is guaranteed. For $T>1$, convergence of the K-SVD updates to a local minimum depends on the accuracy of the sparse-solver used in the sparse coding stage.\cite{aharon06}  

In the K-SVD algorithm, each dictionary update step $i$ sequentially improves both the entries $\mathbf{q}_n\in\mathbf{Q}^i$ and the coefficients in $\mathbf{x}_m\in\mathbf{X}^i$, without change in support. Expressing the coefficients as row vectors $\mathbf{x}_T^n\in\mathbb{R}^N$ and $\mathbf{x}_T^j\in\mathbb{R}^N$, which relate all examples $\mathbf{Y}$ to $\mathbf{q}_n$ and $\mathbf{q}_j$, respectively, the $\ell_2$-penalty from Eq.\ (\ref{eq:dLearnObjective}) is rewritten as 
%
\begin{align}
\begin{split}
\|\mathbf{Y}-\mathbf{Q}\mathbf{X}\|^2_\mathcal{F} &=\bigg\|\mathbf{Y}-\sum_{n=1}^N\mathbf{q}_n\mathbf{x}^n_T\bigg\|^2_\mathcal{F} \\
 &= \|\mathbf{E}_j-\mathbf{q}_j\mathbf{x}^j_T\|^2_\mathcal{F},
\label{eq:ksvdSeparate}
\end{split}
\end{align}
%
where
%
\begin{equation}
\mathbf{E}_j = \bigg{(}\mathbf{Y}-\sum_{n\ne j}\mathbf{q}_n\mathbf{x}^n_T\bigg{)}.
\label{eq:ksvdSeparate2}
\end{equation}
%
Thus, in Eq.\ (\ref{eq:ksvdSeparate}) the $\ell_2$-penalty is separated into an error term $\mathbf{E}_j=[\mathbf{e}_{j,1},...,\mathbf{e}_{j,M}]\in\mathbb{R}^{K\times M}$, which is the error for all examples $\mathbf{Y}$ if $\mathbf{q}_j$ is excluded from their reconstruction, and the product of the excluded entry $\mathbf{q}_j$ and coefficients $\mathbf{x}_T^j\in\mathbb{R}^N$.

An update to the dictionary entry $\mathbf{q}_j$ and coefficients $\mathbf{x}_T^j$ which minimizes Eq.\ (\ref{eq:ksvdSeparate}) is found by taking the SVD of $\mathbf{E}_j$, which provides the best rank-1 approximation of $\mathbf{E}_j$. However, many of the entries in $\mathbf{x}_T^j$ are zero (corresponding to examples which don't use $\mathbf{q}_j$). To properly update $\mathbf{q}_j$ and $\mathbf{x}_T^j$ with SVD, Eq.\ (\ref{eq:ksvdSeparate}) must be restricted to examples $\mathbf{y}_m$ which use $\mathbf{q}_j$
%
\begin{equation}
\|\mathbf{E}_j^R-\mathbf{q}_j\mathbf{x}^j_R\|^2_\mathcal{F},
\label{eq:restricted}
\end{equation}
%
where $\mathbf{E}_j^R$ and $\mathbf{x}_R^j$ are entries in $\mathbf{E}_j$ and $\mathbf{x}_T^j$, respectively, corresponding to examples $\mathbf{y}_m$ which use $\mathbf{q}_j$, and are defined as
%
\begin{align}
\begin{split}
\mathbf{E}_j^R=\big\{\mathbf{e}_{j,l}\big|\forall_l, \ x_l^j \ne 0\big\}, \ 
\mathbf{x}_R^j=\big\{x_l^j\big| \ \forall_l, \ x_l^j \ne 0\big\}.
\label{eq:restricted}
\end{split}
\end{align}
%
Thus for each K-SVD iteration, the dictionary entries and coefficients are sequentially updated as the SVD of $\mathbf{E}^R_j=\mathbf{USV}^{\rm{T}}$. The dictionary entry $\mathbf{q}_j^i$ is updated with the first column in $\mathbf{U}$ and the coefficient vector $\mathbf{x}^j_R$ is updated as the product of the first singular value $\mathbf{S}(1,1)$ with the first column of $\mathbf{V}$. The K-SVD algorithm is given in Table \ref{algo:ksvd}.
\begin{table}[h!]
\begin{center}
\caption{The K-SVD Algorithm (Ref. \ \onlinecite{aharon06})}
\label{algo:ksvd}
\begin{tabular}{ll}
\hline\hline
    & Given: $\mathbf{Y}\in\mathbb{R}^{K\times M}$, $\mathbf{Q}^0\in\mathbb{R}^{K\times N}$, $T\in\mathbb{N}$, and $i=0$ \\ \hline
    & Repeat until convergence: \\
  1.  & \text{Sparse coding}\\
 & \hspace{2ex} \text{for} $m = 1:M$\\
 & \hspace{4ex}  solve Eq.\ (\ref{eq:sparseObject}) using any sparse solver\\
\ \ a:  & \hspace{4ex} $\widehat{\mathbf{x}}_{m}=\underset{\mathbf{x}_m\in\mathbb{R}^N}{\arg\min} \|\mathbf{y}_m-\mathbf{Q}^i\mathbf{x}_m\|_2 \ \ \text{subject to} \ \ \|\mathbf{x}_m\|_0\le T$ \\ 
    & \hspace{2ex} \text{end} \\
\ \ b:  & \hspace{2ex} $\mathbf{X} = [\widehat{\mathbf{x}}_1 ,...,\widehat{\mathbf{x}}_M]$ \\
  2. & \text{Dictionary update}\\
 & \hspace{2ex} for $j = 1:N$\\
\ \ a: & \hspace{4ex} compute reconstruction error $\mathbf{E}_j$ as \\ 
   & \hspace{6ex} $\mathbf{E}_j=\mathbf{Y}-\sum\limits_{n\ne j}\mathbf{q}_n^i\mathbf{x}^n_T$ \\
\ \  b: & \hspace{4ex} obtain $\mathbf{E}_j^R$, $\mathbf{x}_R^j$ corresponding to nonzero $\mathbf{x}_T^j$ \\
\ \  c: & \hspace{4ex} apply SVD to $\mathbf{E}_j^R$ \\
   & \hspace{6ex} $\mathbf{E}_j^R=\mathbf{USV}^{\rm{T}}$ \\
\ \  d: & \hspace{4ex} update $\mathbf{q}_j^{i}$: $\mathbf{q}_j^{i}=\mathbf{U}(:,1)$ \\
\ \  e: & \hspace{4ex} update $\mathbf{x}_R^j$: $\mathbf{x}_R^j=\mathbf{V}(:,1)\mathbf{S}(1,1)$ \\
  & \hspace{2ex} \text{end} \\
\ \ f:   & \hspace{2ex} $\mathbf{Q}^{i+1}=\mathbf{Q}^{i}$ \\
 & \hspace{2ex} $i=i+1$ \\
\hline\hline
\end{tabular}
\end{center}
\end{table}

The dictionary $\mathbf{Q}$ is initialized using $N$ randomly selected, $\ell_2$-normalized examples from $\mathbf{Y}$.\cite{aharon06}\textsuperscript{,}\cite{elad2010} During the iterations, one or more dictionary entries may become unused. If this occurs, the unused entries are replaced using the most poorly represented examples $\mathbf{y}_m$ ($\ell_2$-normlized), determined by reconstruction error.

\section{Experimental results}
\begin{figure}[t]
\begin{tabular}{cc}
\includegraphics[height = 2.6in]{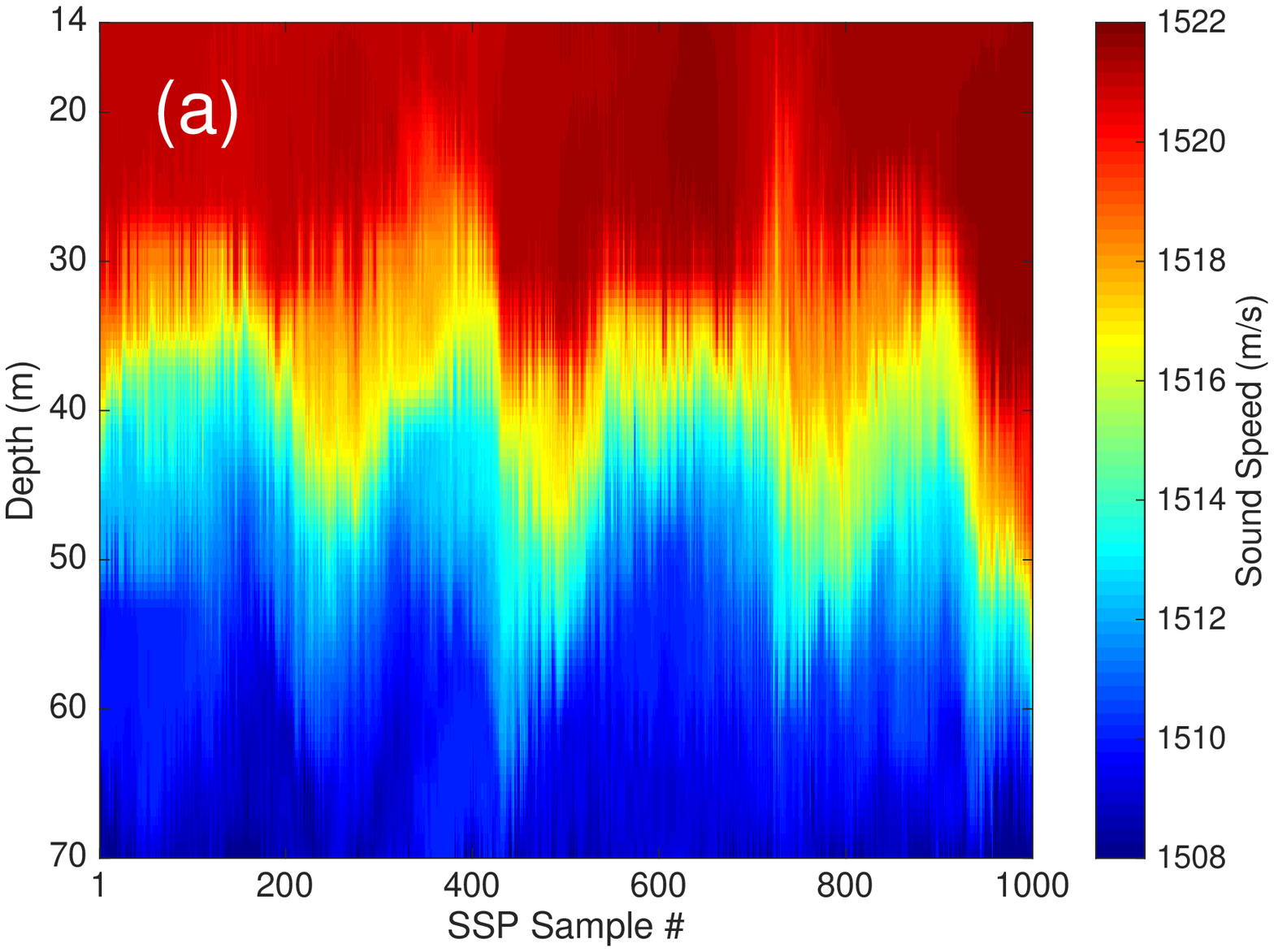} \\
  \includegraphics[height = 2.6in]{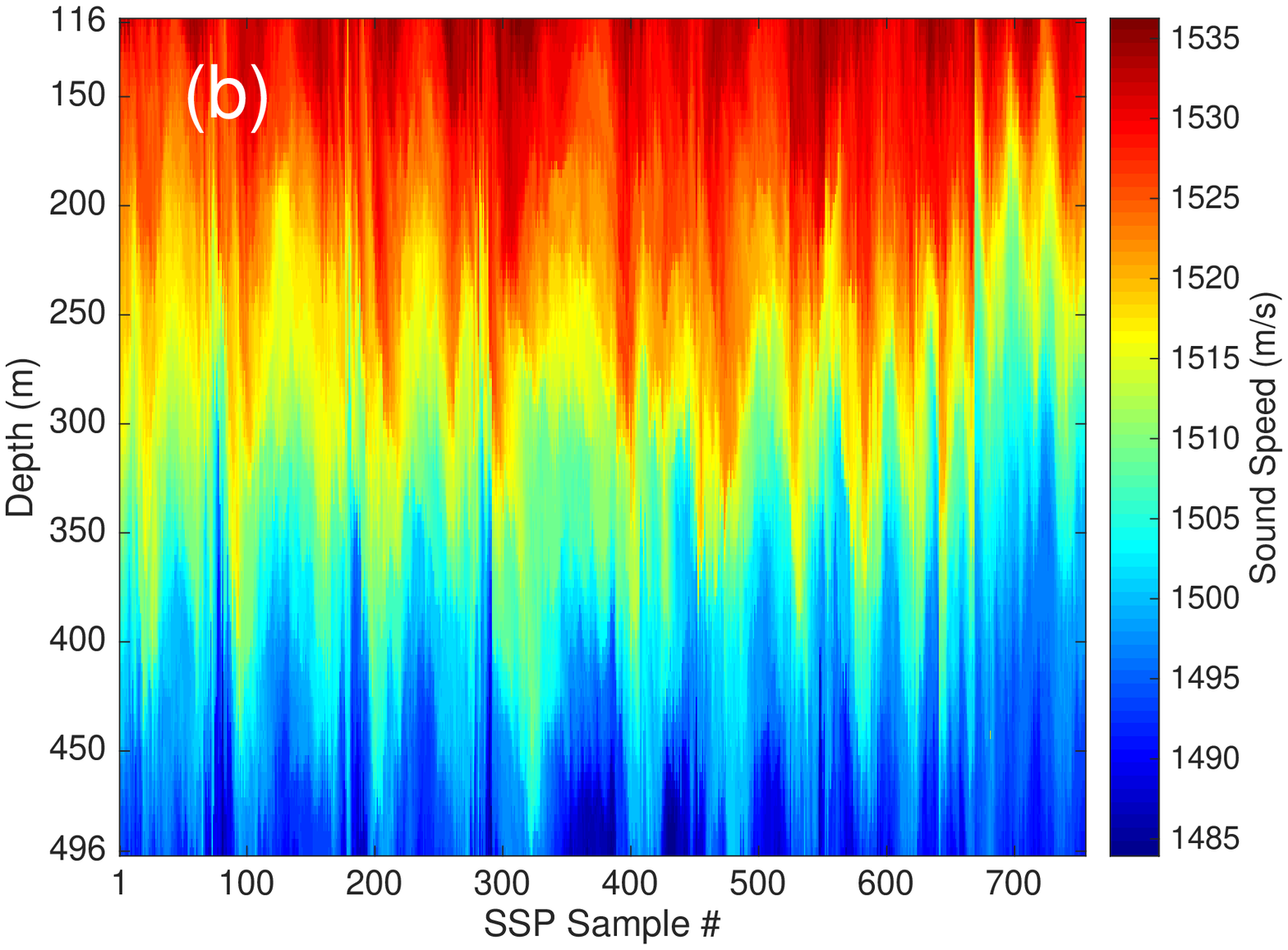}
  \end{tabular}
\caption{(Color online) SSP data from (a) HF-97 experiment and (b) SCS.}
\label{fig:sspHeat}
\end{figure}
\begin{figure*}[t]
\begin{tabular}{cc}
  \includegraphics[height = 2.2in]{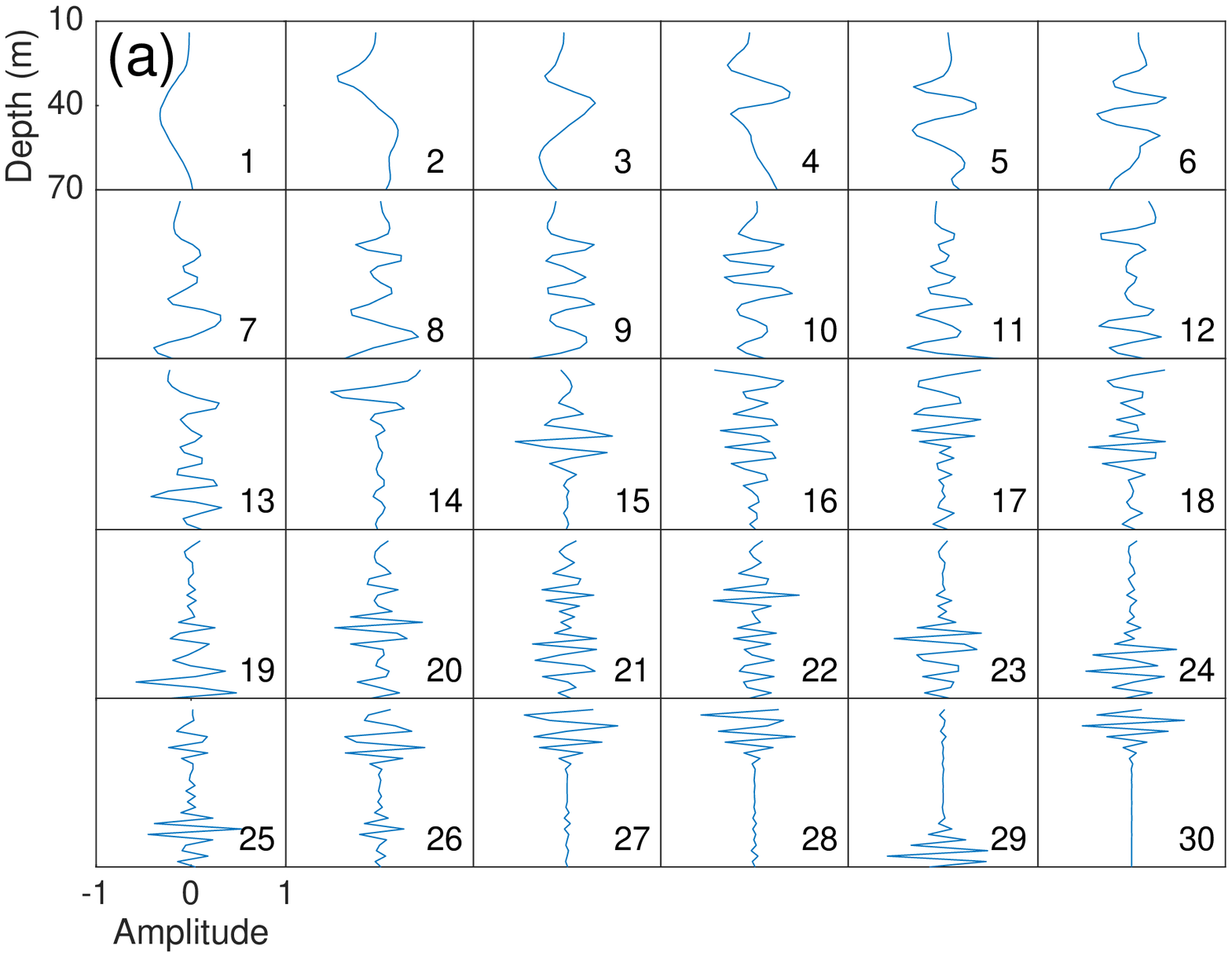} &
  \includegraphics[height = 2.2in]{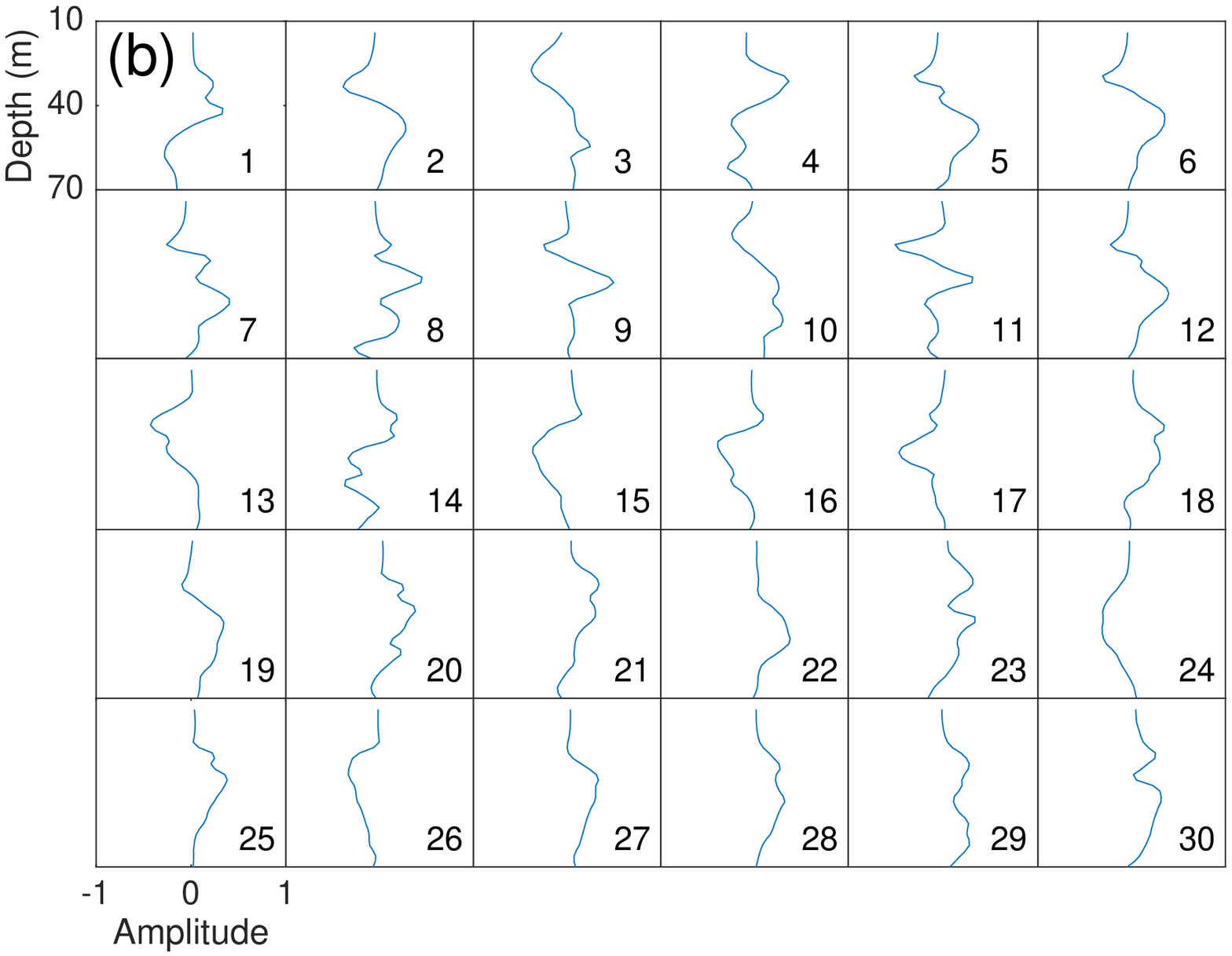} \\
    \includegraphics[height = 1.02in]{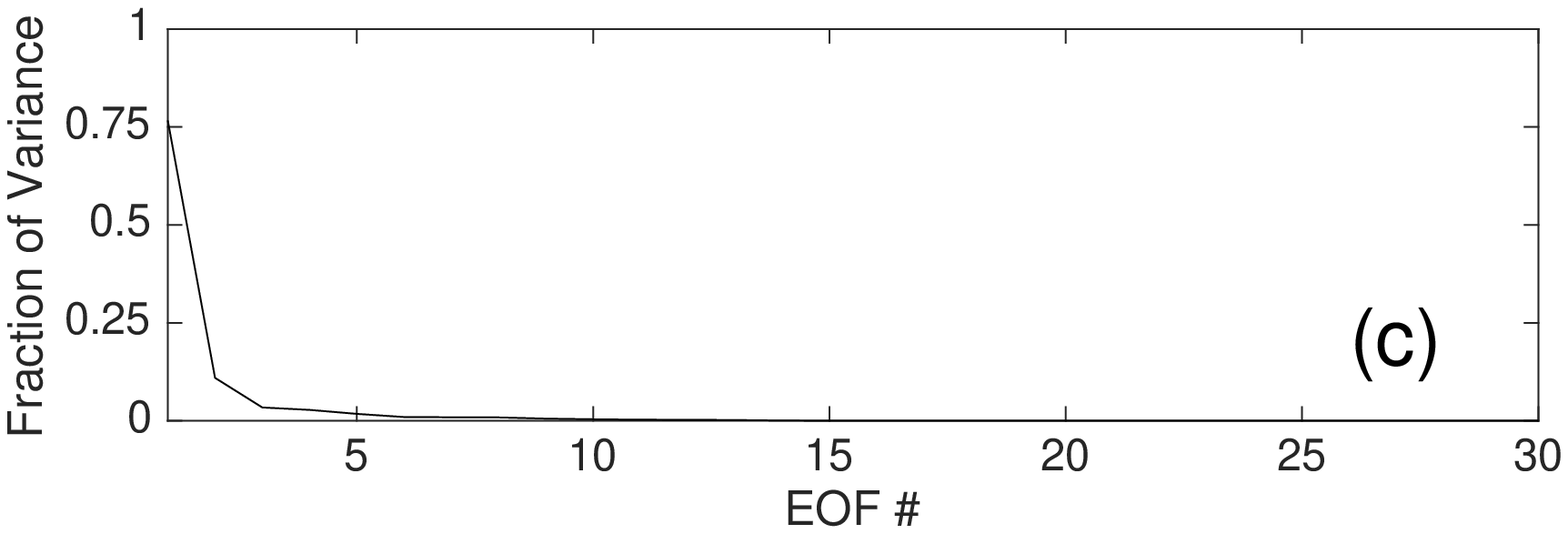} &
  \includegraphics[height = 1.02in]{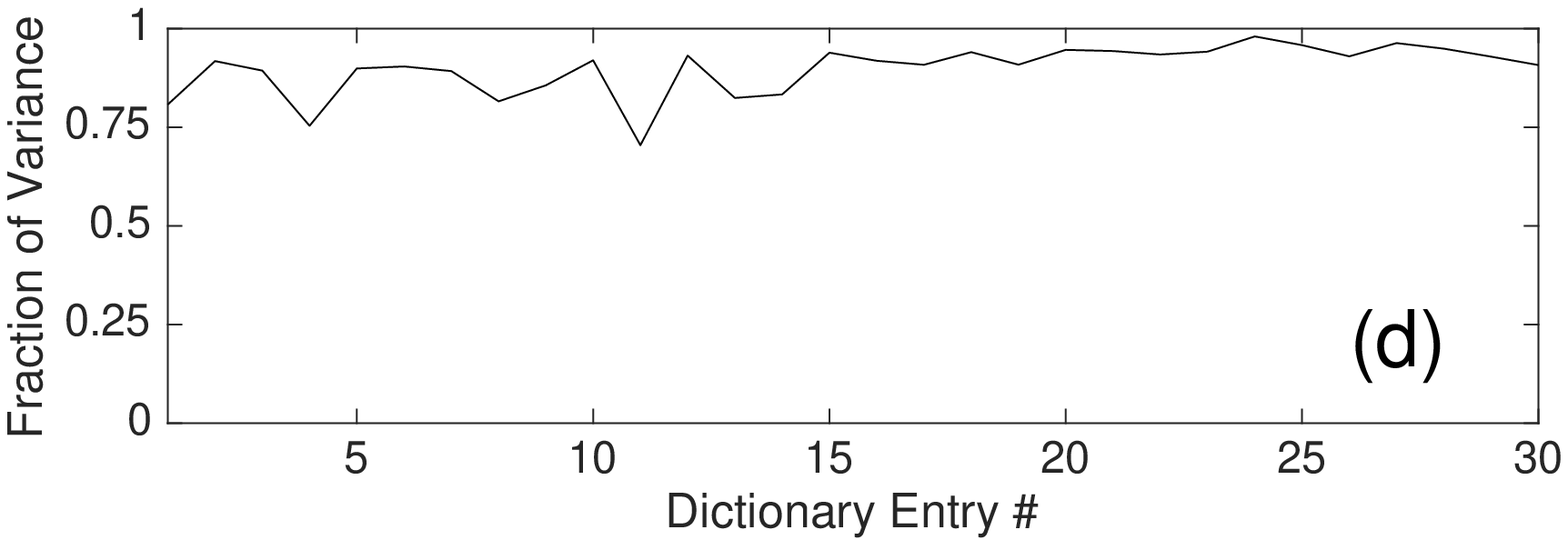} \\
  \includegraphics[height = 2.4in]{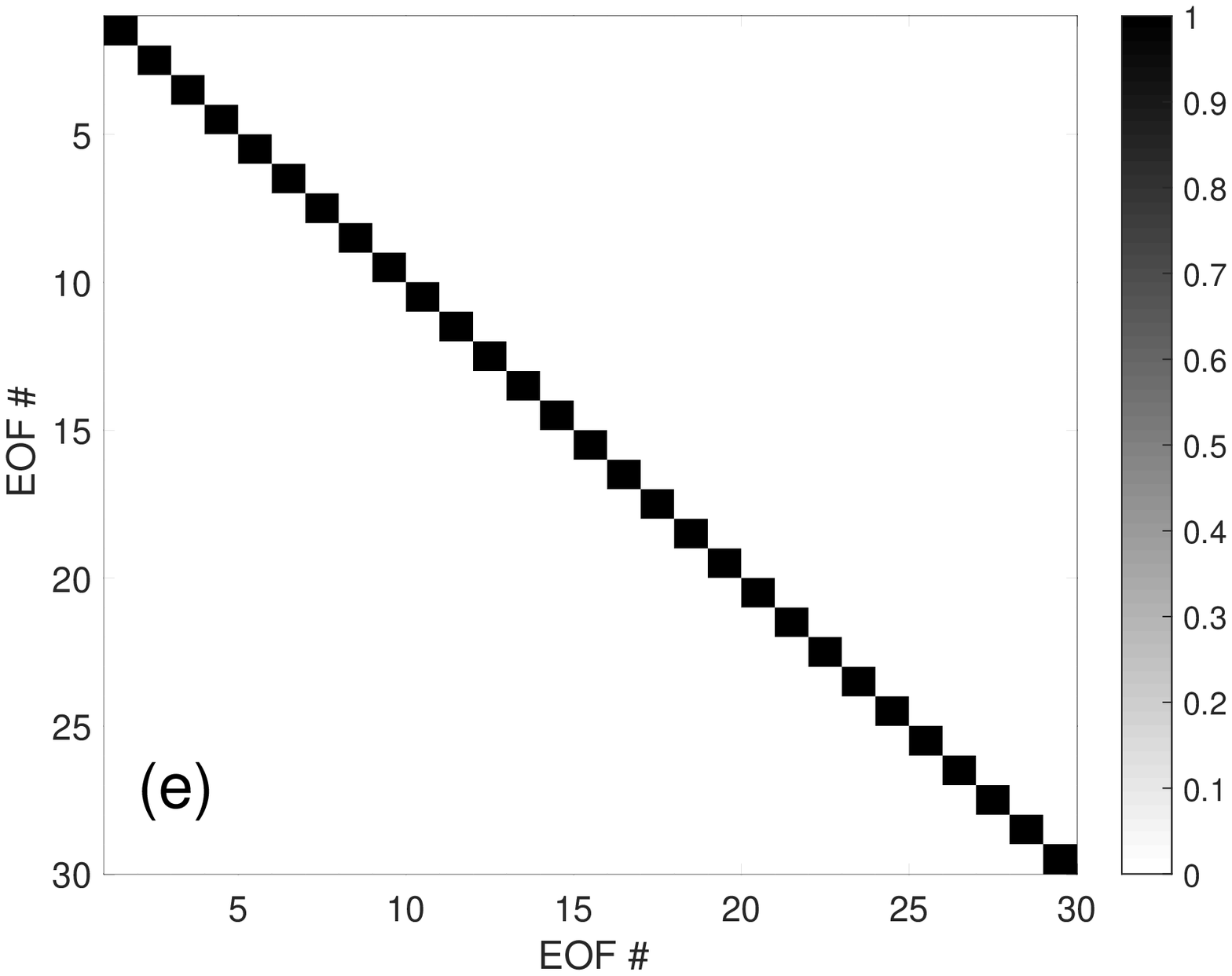} &
  \includegraphics[height = 2.4in]{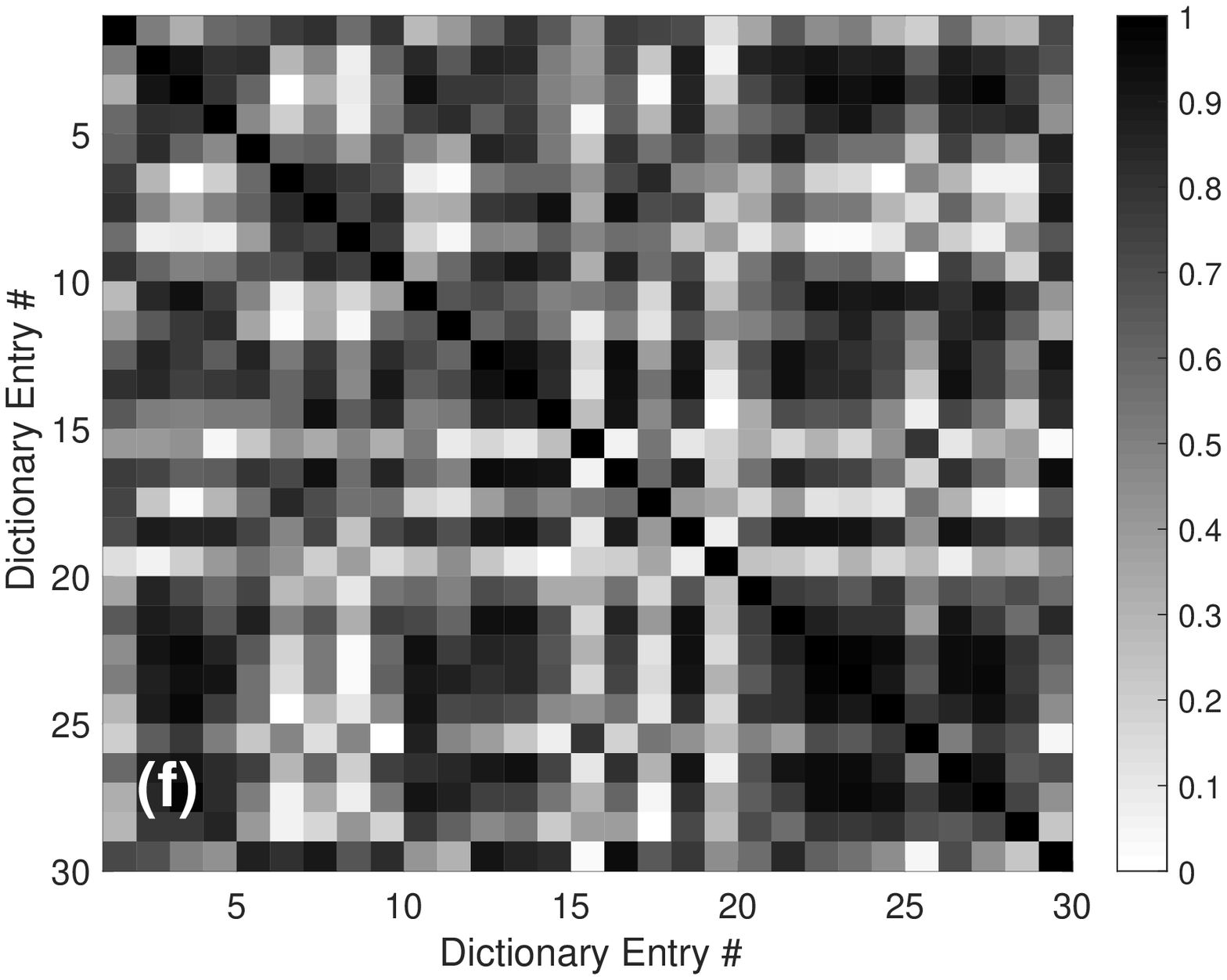} 
    \end{tabular}
\caption{(Color online) HF-97: (a) EOFs and (b) LD entries ($N=K$ and $T=1$, sorted by variance $\sigma_{\mathbf{q}_n}^2$). Fraction of (c) total SSP variance explained by EOFs and (d) SSP variance explained for examples using LD entries. Coherence of (e) EOFs and (f) LD entries.}
\label{fig:eofsVSlds}
\end{figure*}
\begin{figure}
\begin{tabular}{cc}
  \includegraphics[height = 2.5in]{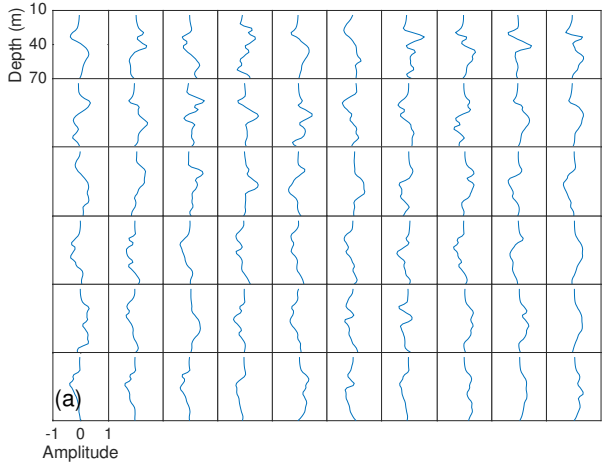} \\
  \includegraphics[height = 2.5in]{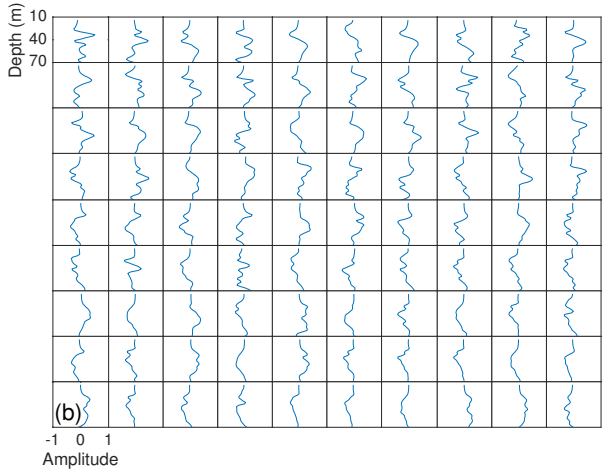} \\
  \includegraphics[height = 2.5in]{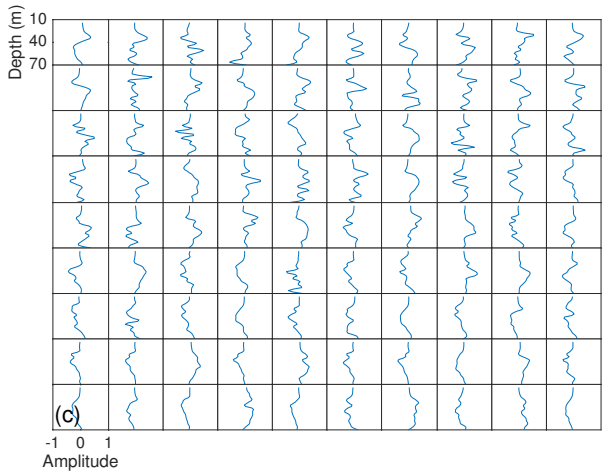}
  \end{tabular}
\caption{(Color online) HF-97: LD entries (a) $N=60$ and $T=1$,  (b) $N=90$ and $T=1$, and (c) $N=90$ and $T=5$. Dictionary entries are sorted in descending variance $\sigma_{\mathbf{q}_n}^2$.}
\label{fig:lds_hf97}
\end{figure}
\begin{figure}
\begin{tabular}{cc}
  \includegraphics[height = 2.5in]{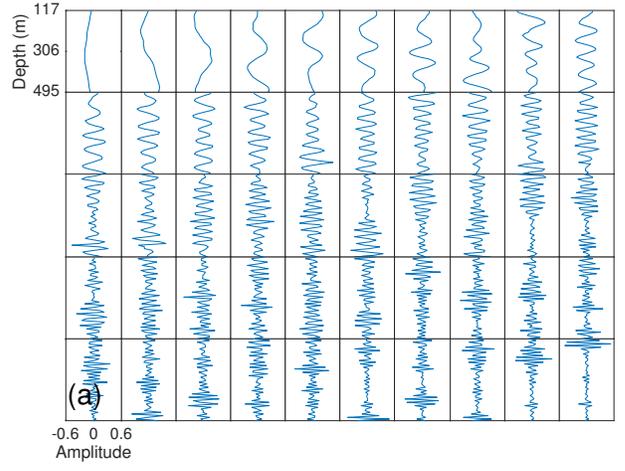} \\
  \includegraphics[height = 2.5in]{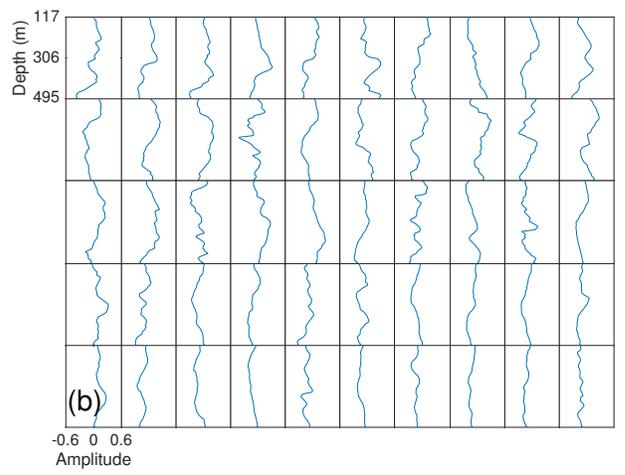} \\
  \includegraphics[height = 2.5in]{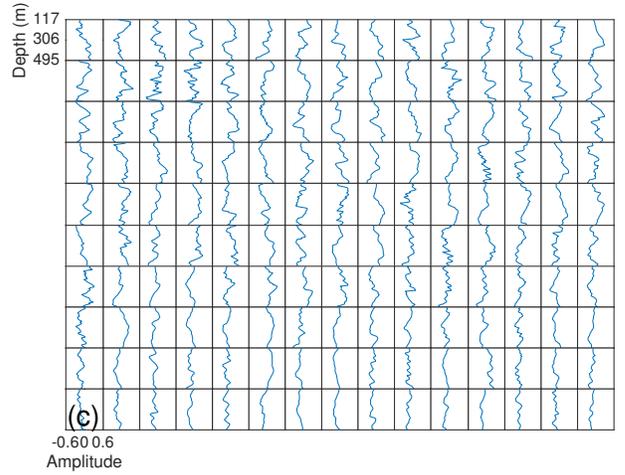}
  \end{tabular}
\caption{(Color online) SCS: EOFs (a) and LD entries; (b) $N=K=50$ and $T=1$ and (c) $N=150$ and $T=1$. Dictionary entries are sorted in descending variance $\sigma_{\mathbf{q}_n}^2$.}
\label{fig:lds_soChina}
\end{figure}
\begin{figure}
\hspace*{-.35in}  
 \includegraphics[width = 4.3in]{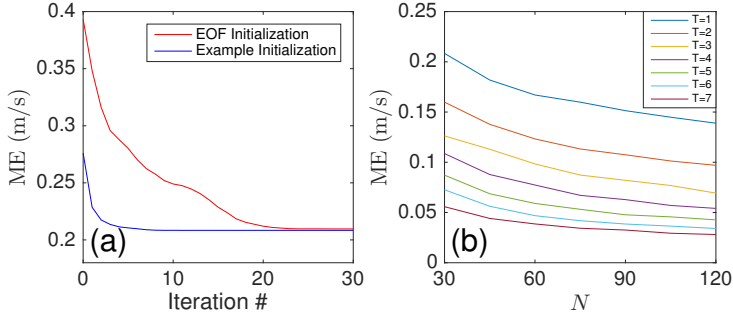}
\caption{(Color online) HF-97: (a) Convergence of LD ($N=30$, $T=1$) mean reconstruction error (ME), initialized using EOFs or $N$ randomly selected examples from $\mathbf{Y}$. (b) ME versus non-zero coefficients $T$ and number of dictionary entries $N$.}
\label{fig:ini_dSize}
\end{figure}
\begin{figure}
\hspace*{-.2in}  
  \includegraphics[width = 3.8in]{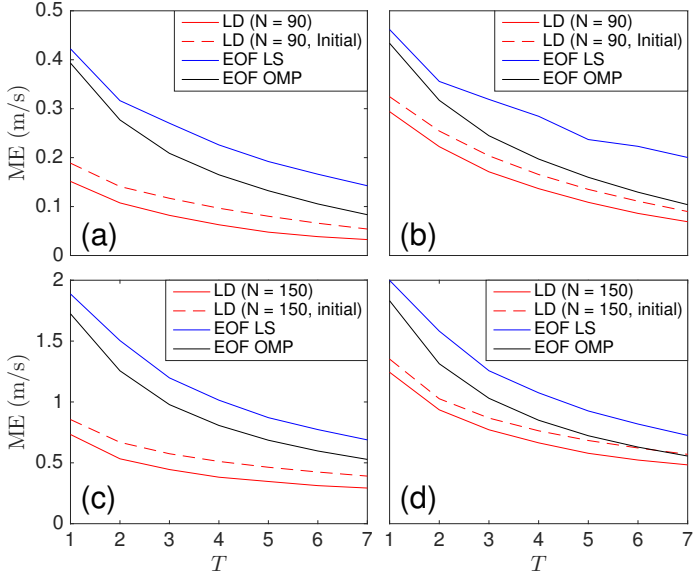}
\caption{(Color online) Mean reconstruction error (ME) versus $T$ using  EOFs (solved using LS and OMP) and LDs ($N=90$ for HF-97 and $N=150$ for SCS) for (a) HF- 97 and (c) SCS. Mean reconstruction error $\rm{ME}_{CV}$  for out-of-sample data calculated with K-fold cross validation for J = 10 folds, (b) HF-97 and (d) SCS.}
\label{fig:error_vs_sparsity}
\end{figure}
\begin{figure}[t]
\begin{tabular}{cc}
  \includegraphics[height = 2.5in]{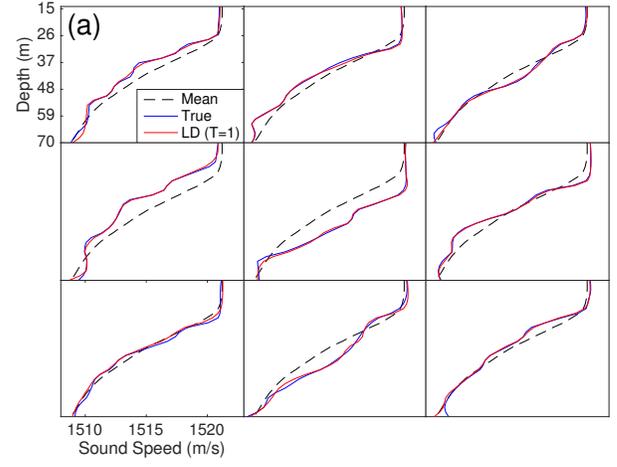} \\
  \includegraphics[height = 2.5in]{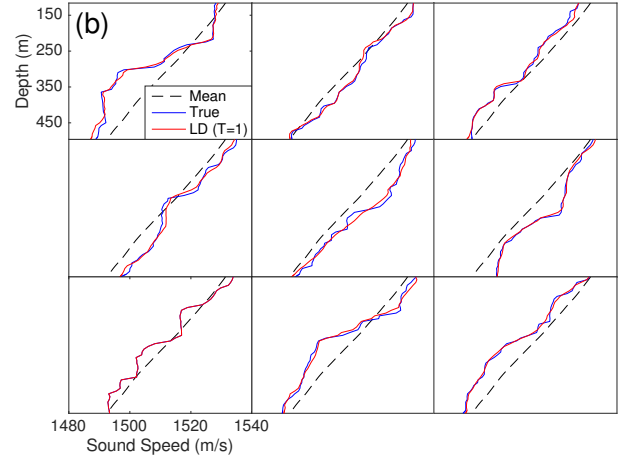}
  \end{tabular}
\caption{(Color online) True SSP reconstruction of 9 example profiles using one coefficient ($T=1$) from LD for (a) HF-97 ($N=90$) and (b) SCS ($N=150$).}
\label{fig:sspEst_examps}
\end{figure}
\begin{figure}
  \includegraphics[height = 2.5in]{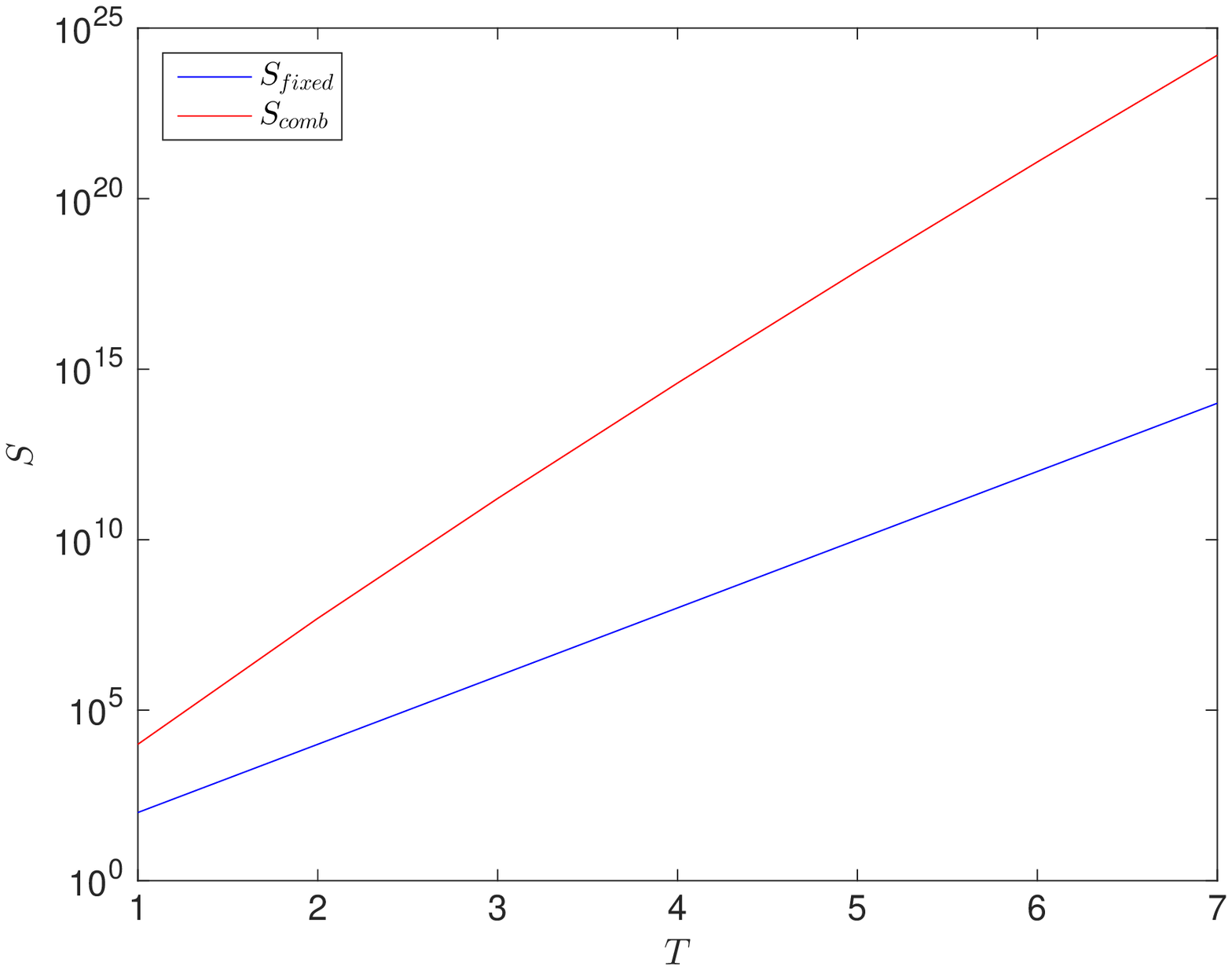}
\caption{(Color online) Number of candidate solutions $S$ for SSP inversion versus $T$, $S_{fixed}$ using fixed indices and $S_{comb}$ best combination of coefficients. Each coefficient is discretized with $H=100$ for dictionary $\mathbf{Q}\in\mathbb{R}^{K\times N}$ with $N=100$.}
\label{fig:searchSize}
\end{figure}
To demonstrate the usefulness of the dictionary learning approach, we here analyze two data sets: (1) thermistor data from the HF-97 acoustics experiment,\cite{carbone2000}\textsuperscript{,}\cite{hodgkiss2002} conducted off the coast of Point Loma, CA and (2) conductivity, temperature, and depth (CTD) data collected across the Luzon Strait near the South China Sea (SCS).\cite{pinkel} Training data $\mathbf{Y}$ were derived from the data sets by converting raw thermistor and CTD data to SSPs and subtracting the mean. The HF-97 thermistor data was recorded every 15 s, over a 48 hour period, from 14 to 70 m depth, with 4 m spacing (15 points). The full 11,488 profile data set was down-sampled to $M=1000$ profiles for the training set, and SSPs were interpolated to $K=30$ points using a shape-preserving cubic spline. The SCS CTD data was recorded at about 1 m resolution from 116 to 496 m depth (384 points). From the SCS data set, $M=755$ profiles were used as the training set, and the profiles were uniformly down-sampled to $K=50$ points. The SSP data sets are shown in Fig.\ \ref{fig:sspHeat}. Both data sets have small and large spatiotemporal variations.

EOFs were calculated from the SVD (Eq.\ \ref{eq:pcaAnal}) and LDs (learned dictionaries) were generated with the K-SVD algorithm (Table \ref{algo:ksvd}), using OMP for the sparse coding stage. The number of non-zero coefficients solved with OMP for each dictionary was held fixed at exactly $T$ non-zero coefficients. The initial dictionary $\mathbf{Q}^0$ was populated using randomly selected examples from the training sets $\mathbf{Y}$.

\subsection{Learning SSP dictionaries from data}
Here, LDs and EOFs were generated using the full SSP data from HF-97 ($M=1000$) and SCS ($M=755$). The EOFs and LDs from HF-97 are shown in Fig.\ \ref{fig:eofsVSlds}--\ref{fig:lds_hf97} and from the SCS in Fig.\ \ref{fig:lds_soChina}. The HF-97 LD, with $N=K$ and $T=1$, is compared to the EOFs ($K=30$) in Fig.\ \ref{fig:eofsVSlds}. Only the leading order EOFs (Fig.\ \ref{fig:eofsVSlds}(a)) are informative of ocean SSP variability whereas all shape functions in the LD (Fig.\ \ref{fig:eofsVSlds}(b)) are informative (Fig.\ \ref{fig:eofsVSlds}(c)--(d)). This behavior is also evident for the SCS data set (Fig.\ \ref{fig:lds_soChina}). The EOFs ($K=50$) calculated from the full training set are shown in Fig.\ \ref{fig:lds_soChina}(a), and the LD entries for $N=50$ and $T=1$ sparse coefficient are shown in Fig.\ \ref{fig:lds_soChina}(b). The overcomplete LDs for the HF-97  data shown in Fig.\ \ref{fig:lds_hf97} and for the SCS data in Fig.\ \ref{fig:lds_soChina}(c).

As illustrated in Fig.\ \ref{fig:featureSpace}, by relaxing the requirement of orthogonality for the shape functions, the shape functions can better fit the data and thereby achieve greater compression. The Gram matrix $\mathbf{G}$, which gives the coherence of matrix columns, is defined for a matrix $\mathbf{A}$ with unit $\ell_2$-norm columns as $\mathbf{G}=|\mathbf{A}^{\rm{T}}\mathbf{A}|$. The Gram matrix for the EOFs (Fig.\ \ref{fig:eofsVSlds}(e)) shows the shapes in the EOF dictionary are orthogonal ($\mathbf{G=I}$, by definition), whereas those of the LD (Fig.\ \ref{fig:eofsVSlds}(f)) are not. 

\subsection{Reconstruction of SSP training data}
In this section, EOFs and LDs are trained on the full SSP data sets $\mathbf{Y}=[\mathbf{y}_1,..., \mathbf{y}_M]$. Reconstruction performance of the EOF and LDs are then evaluated on SSPs within the training set, using a mean error metric. 

The coefficients for the learned $\mathbf{Q}$ and initial $\mathbf{Q}^0$ dictionaries $\widehat{\mathbf{x}}_m$ are solved from the sparse objective (Eq.\ (\ref{eq:sparseObject})) using OMP. The least squares (LS) solution for the $T$ leading-order coefficients $\mathbf{x}_L\in\mathbb{R}^{T}$ from the EOFs $\mathbf{P}$ were solved by Eq.\ (\ref{eq:eofPseudo}). The best combination of $T$ EOF coefficients was solved from the sparse objective (Eq.\ (\ref{eq:sparseObject})) using OMP. Given the coefficients $\mathbf{X}=[\mathbf{x}_1,...,\mathbf{x}_m]$ describing examples $\mathbf{Y}=[\mathbf{y}_1,...,\mathbf{y}_m]$, the reconstructed examples $\widehat{\mathbf{Y}}=[\widehat{\mathbf{y}}_1,...,\widehat{\mathbf{y}}_m]$ are given by $\widehat{\mathbf{Y}}=\mathbf{Q}\widehat{\mathbf{X}}$. The mean reconstruction error $\rm{ME}$ for the training set is then
%
\begin{equation}
\text{ME}=\frac{1}{KM}\|\mathbf{Y}-\widehat{\mathbf{Y}}\|_1.
\label{eq:meanErrorTrain}
\end{equation}
%
We here use the $\ell_1$-norm to stress the robustness of the LD reconstruction. 

To illustrate the optimality of LDs for SSP compression, the K-SVD algorithm was run using EOFs as the initial dictionary $\mathbf{Q}^0$ for $T=1$ non-zero coefficient. The convergence of $\text{ME}$ for the K-SVD iterations is shown in Fig.\ \ref{fig:ini_dSize}(a). After 30 K-SVD iterations, the mean error of the $M=1000$ profile training set is decreased by nearly half. The convergence is much faster for $\mathbf{Q}^0$ consisting of randomly selected examples from $\mathbf{Y}$.

For LDs, increasing the number of entries $N$ or increasing the number of sparse coefficients $T$ will always reduce the reconstruction error ($N$ and $T$ are decided with computational considerations). The effect of $N$ and $T$ on the mean reconstruction error for the HF-97 data is shown in Fig.\ \ref{fig:ini_dSize}(b). The errors are calculated for the range $N=K$ to $N=4K$ and the dictionaries were optimized to use a fixed number non-zero coefficients ($T$). 

The reconstruction error using the EOF dictionary is compared to results from LDs $\mathbf{Q}$ with $N=3K$, using $T$ non-zero coefficients. In Fig.\ \ref{fig:error_vs_sparsity}[(a) and (c)] results are shown for the HF-97 ($N=90$) and SCS ($N=150$) data, respectively. Coefficients describing each example $\mathbf{y}_m$, were solved (1) from the LD $\mathbf{Q}$, (2) from $\mathbf{Q}^0$, the dictionary consisting of $N$ randomly chosen examples from the training set (to illustrate improvements in reconstruction error made in the K-SVD iterations), (3) the leading order EOFs, and (4) the best combination of EOFs. The mean SSP reconstruction error using the LDs trained for each sparsity $T$ is less than EOF reconstruction, for either leading order coefficients or best coefficient combination, for all values of $T$ shown. The best combination of EOF coefficients, chosen approximately using OMP, achieves less error than the LS solution to the leading order EOFs, with added cost of search. 

Just one LD entry achieves the same $\rm{ME}$ as more than 6 leading order EOF coefficients, or greater than 4 EOF coefficients chosen by search (Fig.\ \ref{fig:error_vs_sparsity}[(a) and (c)]). To illustrate the representational power of the LD entries, both true and reconstructed SSPs are shown in Fig.\ \ref{fig:sspEst_examps}(a) for the HF-97 data and in Fig.\ \ref{fig:sspEst_examps}(b) for the SCS data. Nine true SSP examples from each training set, for HF-97 (SCS) taken at 100 (80) point intervals from $m=100$ to 900 (80 to 720), are reconstructed using one LD coefficient. It is shown for each case, that nearly all of the SSP variability is captured using a single LD coefficient.

\subsection{Cross-validation of SSP reconstruction}
The out of sample SSP reconstruction performance of LDs and EOFs is tested using K-fold cross-validation.\cite{hastie2009} The entire SSP data set $\mathbf{Y}$ of $M$ profiles, for each experiment, is divided into $J$ subsets with equal numbers of profiles $\mathbf{Y}=[\mathbf{Y}_1,...,\mathbf{Y}_J]$, where the fold $\mathbf{Y}_j\in\mathbb{R}^{K\times (M/J)}$. For each of the $J$ folds: (1) $\mathbf{Y}_j$ is the set of out of sample test cases, and the training set $\mathbf{Y}_{tr}$ is
%
\begin{equation}
\mathbf{Y}_{tr}=\big\{\mathbf{Y}_l\big| \ \forall_{l\ne j}\big\};
\label{eq:restricted}
\end{equation}
%
(2) the LD $\mathbf{Q}_j$ and EOFs are derived using $\mathbf{Y}_{tr}$; and (3) coefficients estimating test samples $\mathbf{Y}_j$ are solved for $\mathbf{Q}_j$ with sparse processor Eq.\ (\ref{eq:sparseObject}), and for EOFs by solving for leading order terms and by solving with sparse processor. The out of sample error from cross validation $\rm{ME}_{CV}$ for each method is then
%
\begin{equation}
\text{ME}_{CV}=\frac{1}{KM}\sum_{j=1}^J\|\mathbf{Y}_j-\widehat{\mathbf{Y}}_j\|_1.
\label{eq:kFolds}
\end{equation}
%
\par
The out of sample reconstruction error $\rm{ME}_{CV}$ increases over the within-training-set estimates for both the learned and EOF dictionaries, as shown in Fig.\ \ref{fig:error_vs_sparsity}[(b) and (d)] for $J=10$ folds. The mean reconstruction error using the LDs, as in the within-training-set estimates, is less than the EOF dictionaries. For both the HF-97 (SCS) data, more than 2 (2) EOF coefficients, choosing best combination by search, or more than 3 (equal to 3) leading-order EOF coefficients solved with LS, are required to achieve the same out of sample performance as one LD entry.

\subsection{Solution space for SSP inversion}
Acoustic inversion for ocean SSP is a non-linear problem. One approach is coefficient search using genetic algorithms.\cite{gerstoft94} Discretizing each coefficient into $H$ values, the number of candidate solutions for $T$ fixed coefficients indices is
%
\begin{equation}
S_\text{fixed}=H^T.
\label{eq:restricted}
\end{equation}
%
If the coefficient indices for the solution can vary, as per dictionary learning with LD $\mathbf{Q}\in\mathbb{R}^{K\times N}$, the number of candidate solutions  $S_\text{comb}$ is
%
\begin{equation}
S_\text{comb}=H^T\frac{N!}{T!(N-T)!}.
\label{eq:restricted}
\end{equation}
%
Using a typical $H=100$ point discretization of the coefficients, the number of possible solutions for fixed and combinatorial dictionary indices are plotted in Fig.\ \ref{fig:searchSize}. Assuming an unknown SSP similar to the training set, the SSP may be constructed up to acceptable resolution using one coefficient from the LD ($10^4$ possible solutions, see Fig.\ \ref{fig:searchSize}). To achieve the similar ME, 7 EOFs coefficients are required ($10^{14}$ possible solutions, Fig.\ \ref{fig:searchSize}) using fixed indices and the best EOF combination requires 5 EOFs ($10^{17}$ possible solutions, Fig.\ \ref{fig:searchSize}).

\section{Conclusion}
Given sufficient training data, dictionary learning generates optimal dictionaries for sparse reconstruction of a given signal class. Since these LDs are not constrained to be orthogonal, the entries fit the distribution of the data such that signal example is approximated using few LD entries. Relative to EOFs, each LD entry is informative to the signal variability.

The K-SVD dictionary learning algorithm is applied to ocean SSP data from the HF-97 and SCS experiments. It is shown that the LDs generated describe ocean SSP variability with high resolution using fewer coefficients than EOFs. As few as one coefficient from a LD describes nearly all the variability in each of the observed ocean SSPs. This performance gain is achieved by the larger number of informative elements in the LDs over EOF dictionaries. Provided sufficient SSP training data is available, LDs can improve SSP inversion resolution with negligible computational expense. This could provide improvements to geoacoustic inversion,\cite{gerstoft94} matched field processing,\cite{bag93}\textsuperscript{,}\cite{verlinden15} and underwater communications.\cite{carbone2000}

\begin{acknowledgments}
\noindent
The authors would like to thank Dr. Robert Pinkel for the use of the South China Sea CTD data. This work is supported by the Office of Naval Research, Grant No. N00014-11-1-0439. 
\end{acknowledgments}


\end{document}